\documentclass[conference,10pt]{IEEEtran}
\usepackage{epsfig,rotating,setspace,latexsym,amsmath,epsf,amssymb,amsfonts,bm,theorem,epstopdf,cite,mathtools,caption,subcaption,enumerate,longtable,accents,bbm,algorithm,graphicx,epsf,authblk,url,color,multirow,comment,tabularx,dsfont}
\usepackage[inline]{enumitem}
\usepackage[noend]{algpseudocode}

\algrenewcommand\algorithmicforall{\textbf{foreach}}
\algrenewcommand\algorithmicindent{.8em}

\newtheorem{theorem}{Theorem}

\newtheorem{corollary}{Corollary}
\newtheorem{definition}{Definition}

\newtheorem{lemma}{Lemma}

\newenvironment{Proof}[1]{\medskip\par\noindent{\bf Proof:\,}\,#1}{{\mbox{\,$\blacksquare$}\par}}
\newcommand{\figref}[1]{\figurename~\ref{#1}}
\IEEEoverridecommandlockouts
\allowdisplaybreaks

\begin{document}

\title{PoW Security-Latency under Random Delays and the Effect of Transaction Fees}

\author[1]{Mustafa Doger}
\author[1]{Sennur Ulukus}
\author[2]{Nail Akar}

\affil[1]{\normalsize University of Maryland, College Park, MD, USA}
\affil[2]{\normalsize Bilkent University, Ankara, T\"{u}rkiye}

\maketitle
\let\thefootnote\relax\footnotetext{This work is done when N.~Akar is on sabbatical leave as a visiting professor at University of Maryland, MD, USA, which is supported in part by the Scientific and Technological Research Council of T\"{u}rkiye  (T\"{u}bitak) 2219-International Postdoctoral Research Fellowship Program.}

\begin{abstract}
    Safety guarantees and security-latency problem of Nakamoto consensus have been extensively studied in the last decade with a bounded delay model. Recent studies have shown that PoW protocol is secure under random delay models as well. In this paper, we analyze the security-latency problem, i.e., how secure a block is, after it becomes $k$-deep in the blockchain, under general random delay distributions. We provide tight and explicit bounds which only require determining the distribution of the number of Poisson arrivals during the random delay. We further consider potential effects of recent Bitcoin halving on the security-latency problem by extending our results.
\end{abstract}

\section{Introduction}

Nakamoto consensus \cite{btc-whitepaper} relies on Proof-of-Work (PoW) and the longest chain protocol to achieve consensus among  distributed nodes. However, adversarial activities and network delays among the distributed nodes pose a threat to the consensus mechanism since they can result in the violation of an agreement made on a specific object, e.g., transaction.

Although most studied models for PoW blockchains assume a bounded network delay on the basis of which they provide safety guarantees, \cite{pow-under-random-safe} showed that PoW longest chain protocols have good security guarantees even when delays are sporadically large and possibly unbounded. In this context, the relation between the amount of time one waits before confirming a transaction and the safety guarantee is referred to as the \emph{security-latency problem}. The security-latency problem was studied extensively for the bounded delay model in \cite{nakamoto-always-wins, guo-close-sec-lat, guo-btc-sec-lat, our-sec-lat-isit,our-sec-lat-extended, cao2023tradeoff, gazi-sec-lat}. Recently, \cite{our-queue-sec-ext-version} extended the security-latency analysis to the case where network delays are i.i.d. and exponentially distributed, and connected the results to the transaction capacity a blockchain can sustain, using batch service queue stability conditions provided in \cite{quan-li-queue-blockchain}.

In this paper, we first consider the security-latency problem, i.e., how secure a transaction is after it becomes $k$-deep in the longest chain where network delays are random. We provide generalized safety guarantees and conditions under which the transactions are safe. The safety guarantees we provide are extremely simple to evaluate; they involve finite sums of convolutions and only require one to know the distribution of the number of adversarial and honest blocks mined during the random delay interval.

Next, we consider the effect of halving mining rewards in blockchain systems like Bitcoin (BTC) which makes miners increasingly sensitive on transaction fees. The recent halving of BTC in April 2024 also halved the expected revenues of miners per hash/day from March to May 2024 \cite{hashrateindex}. As transaction fees become the main source of compensation for miners' hardware, maintenance and utility costs, for each block cycle, miners may spend some idle time until a certain number of high-paying transactions are accumulated in the mempool, in order to cover their costs. However, idle times spent by honest miners have a negative impact on safety guarantees. This problem can be analyzed with the methods we provide for random delays with some minor modifications. The method we propose in this paper can also be used as a tool to determine various aspects of blockchains such as transaction fees, maintenance cost sustainability and their effects on safety.

\section{System Model}

We abstract out the semantic details of PoW and the longest chain protocols of blockchain systems and simply assume exponential inter-arrival times for blocks whenever miners are active. Each mined block has a maximum size of $b$ transactions and miners observe a mempool of transactions where two types of transactions arrive at the mempools. The first type of transactions have arrival rate $\lambda_h$ and reward the miners a high-fee, thus they are called high priority (HP) transactions. After at least $b_0$ pending HP transactions in the mempool are added to the new block to be mined, the low fee transactions fill the rest to the maximum size of $b$. For the sake of simplicity, we assume $\lambda_h$ is small enough so that the number of HP transactions waiting in the mempool is smaller than $b$ at any time. We explain the implications of this assumption on our results later on. The total mining power in this system has the rate $\mu_m$ and $\beta$ fraction of it is controlled by a single adversarial entity (fully-coordinating adversaries). 

Although honest miners follow the longest chain protocol, each honest miner spends an idle time until there are at least $b_0$ HP transactions in their mempool. Moreover, each block mined by an honest node at height $h$ experiences a network delay of $t_{hi}\in [0,\Delta_h]$ before it becomes available to any other honest node $i$, where $\Delta_h, h\geq 1$ are i.i.d. The adversary is allowed to deviate from the longest chain protocol and does not have to wait for $b_0$ HP transactions (it can simply move its own funds). Further, it controls network delays for each honest block at height $h$ as long as $t_{hi}\in [0,\Delta_h]$, and ties are broken in the adversary's favor. A transaction in this model is said to be confirmed according to the $k$-block confirmation rule, i.e., if it is part of the longest chain of the honest view and there are at least $k-1$ blocks mined on top of it. 

In this paper, we are interested in finding the safety violation probability for transactions under various scenarios. The safety violation, or double spending, refers to the event that a transaction $tx$ is confirmed according to the $k$-block confirmation rule in an honest view and later gets replaced (or discarded) with another conflicting transaction $tx'$. We first provide theoretical bounds for the safety violation probability under the random network delay model without the HP transaction condition. Next, we consider the same model with the HP transaction condition. Our model and results can be applied to various random network delay scenarios including as a sub-case the bounded delay model used extensively in literature \cite{nakamoto-always-wins, guo-close-sec-lat, guo-btc-sec-lat, our-sec-lat-isit,our-sec-lat-extended, cao2023tradeoff, gazi-sec-lat,our-queue-sec-ext-version}.

\section{Upper Bounds}

We assume that a transaction $tx$ arrives at the mempools of miners at time $\tau_0$ (which is not at the control of the adversary) to be included in the next honest block. As it is done in \cite{guo-close-sec-lat,guo-btc-sec-lat,our-sec-lat-isit,our-queue-sec-ext-version,cao2023tradeoff}, we divide our analysis into three phases: pre-mining gain, confirmation interval and post confirmation race. Pre-mining gain refers to the adversarial lead accumulated before $tx$ arrives at the system where the adversary tries to build a private chain that contains a conflicting transaction $tx'$ to gain some lead over the honest chain. Hence, the lead refers to the height difference between the longest chain and the longest honest chain at $\tau_0$. Letting $\tau_k$ denote the first time when the block containing $tx$ is $k$-deep in all honest views and there are at least $b_0$ HP transactions in all mempools, the confirmation interval spans $[\tau_0,\tau_k]$. The post confirmation race refers to the race between the chains that contain $tx$ and $tx'$ after $\tau_k$.
 
To keep things simple, we first assume $b_0=0$ and re-analyze the model provided in \cite{our-queue-sec-ext-version} in a generalized manner for different distributions and tighten the bounds. Next, to introduce the effect of $b_0$ on the adversarial strategies, we consider different sample paths of block arrivals and HP transaction arrivals. Then, we argue a model for the adversarial strategy to upper bound the safety violation probability.
 
\subsection{The Case of $b_0=0$}
We first provide some definitions to analyze the race. Let $L_t$ denote the lead at time $t$ and $A_{r,t}$ denote the number of adversarial blocks mined during the interval $(r,t]$. 

\begin{definition}
     {\normalfont \textbf{(Pacer Process)}} Starting from $r\geq0$, the pacer process $P^{r}$ is created as follows: The first honest block that arrives after $r$ is mined on a new height $h$ and delayed by maximally allowed $\Delta_h$ (all other honest blocks during this delay are also delayed maximally and create forks). This block is called the first pacer of $P^r$. After the first pacer is published, the next honest block is going to be mined on a new height $h+1$ and the same delay strategy is applied to create the second pacer and so on. The number of pacers mined during the interval $(s,t]$ associated with $P^r$ is denoted by $P^r_{s,t}$
\end{definition}

\begin{definition}
     {\normalfont \textbf{(Jumper)}} Jumpers are the first honest blocks mined on their heights \cite{cao2023tradeoff}. We let $J_{r,t}$ denote the number of jumpers mined in the interval $(r,t]$.
\end{definition}

Note that $L_t\geq0$, since the adversary can always choose to mine on the longest honest chain to extend its lead. Further, the lead of the adversary increases by at most one when the adversary mines a block. It increases by exactly one if the block is mined on top of the longest chain and kept private. Before $\tau_0$, whenever an honest jumper is mined, the adversarial lead drops by at least one unless the lead is already zero. If adversary always keeps the blocks it mines private, then each jumper decreases the lead by exactly one unless it is already zero. These imply the following \cite{cao2023tradeoff}
\begin{align}
        L_{t}&\leq\sup_{s\in[0,t]}\{A_{s,t}-J_{s,t}\}. \label{raw_lead_bound}
\end{align}

\begin{lemma}\label{jumper-start-maximizes}
    Let $\mathcal{T}_J$ denote the set of mining times of jumper blocks in the interval $[0,t]$. Then, 
    \begin{align}
        \sup_{s\in[0,t]}\{A_{s,t}-J_{s,t}\}=\sup_{s\in\mathcal{T}_J}\{A_{s,t}-J_{s,t}\}.
    \end{align}
\end{lemma}

\begin{Proof}
    Clearly, $\sup_{s\in[0,t]}\{A_{s,t}-J_{s,t}\}\geq\sup_{s\in\mathcal{T}_J}\{A_{s,t}-J_{s,t}\}$ since if $s\in\mathcal{T}_J$, then $s\in[0,t]$. To prove the other direction, consider any $r^*=\arg\max_{s\in[0,t]}\{A_{s,t}-J_{s,t}\}$. Pick largest $r_{j*}\in\mathcal{T}_J$ with $r_{j*}\leq r^*$. Clearly, $J_{r_{j*},r^*}=0$ by definition and $A_{r_{j*},r^*}\geq0$ which implies $\sup_{s\in[0,t]}\{A_{s,t}-J_{s,t}\}\leq\sup_{s\in\mathcal{T}_J}\{A_{s,t}-J_{s,t}\}$.
\end{Proof}

For a given $t$, honest mining times $\mathcal{T}_H$ and a jumper block mined at $t_{j}\in\mathcal{T}_H$ with height $h_j$, let $J^*_{t_{j},t}$ denote a jumper process associated with delaying the publication of all honest blocks at each height $h_i\geq h_j$ by maximally allowed time $\Delta_{h_i}$. Let $J_{t_{j},t}$ denote the jumper process associated with any other delay strategy for the same honest mining times $\mathcal{T}_H$ starting from jumper block mined at $t_{j}\in\mathcal{T}_H$. It is straightforward to see $J^*_{t_{j},t} \leq J_{t_{j},t}$ since delaying the publications by maximally allowed time results in forkers instead of jumpers. Together with Lemma \ref{jumper-start-maximizes} this implies,
\begin{align}
    L_{t}
    &\leq\sup_{r\in\mathcal{T}_J}\{A_{r,t}-J^*_{r,t}\}.
\end{align}

In other words, the lead is bounded by the adversarial strategy that delays all honest blocks by maximally allowed time after picking a specific honest block as a jumper, which we call starting jumper block. Clearly, as the adversary cannot know the future while making the decision of picking the starting jumper block, the bound we provide is extremely pessimistic since it assumes that the adversary picks the best possible starting jumper block without knowing the future.

Let $C_\Delta$ denote the random variable that represents the number of adversarial blocks mined during a delay interval $\Delta_h$ independent from the process $A$. Let $\Tilde{P}^r$ denote an independent pacer process with same statistics as $P^r$ and $\Tilde{A}$ denote an independent process with the same statistics as $A$. First, note that $A_{t_{j},t_{j}+\Delta_{h_{j}}}$ has the same distribution as $C_\Delta$. Further, note that $J^*_{t_{j},t_{j}+\Delta_{h_{j}}}=0$ and $J^*_{t_{j}+\Delta_{h_{j}},t}=P^{t_{j}+\Delta_{h_{j}}}_{t_{j}+\Delta_{h_{j}},t}$ by definition. Next, $P^{t_{j}+\Delta_{h_{j}}}_{t_{j}+\Delta_{h_{j}},t}$ is i.i.d.~as $\Tilde{P}^0_{0,t-t_{j}-\Delta_{h_{j}}}$ and $A_{t_{j}+\Delta_{h_{j}},t}$ is i.i.d.~as $\Tilde{A}_{0,t-t_{j}-\Delta_{h_{j}}}$. Then, we have,
\begin{align}
    P(L_{t}\geq x)&\leq P(\sup_{r\in\mathcal{T}_J}\{A_{r,t}-J^*_{r,t}\} \geq x)\\
    &\leq P(\sup_{t\geq q\geq0}\{C_\Delta + \Tilde{A}_{0,q}+\Tilde{P}^0_{0,q}\} \geq x). \label{lead_bound_tildes}
\end{align}

\begin{lemma}\label{pacer-end-maximizes}
    Let $\mathcal{T}_{P^0}$ denote the set of mining times of pacer blocks in interval $(0,t]$ associated with $P^0$. Let $s^-$ denote the time right before time $s$. Then, 
    \begin{align}
        \sup_{s\in[0,t]}\{A_{0,s}-P^0_{0,s}\}=\sup_{s\in\mathcal{T}_{P^0}}\{A_{0,s^-}-P^0_{0,s^-}\}.
    \end{align}
\end{lemma}

\begin{Proof}
     If $s=0$, both sides are zero. If $s\in\mathcal{T}_{P^0}$, then $s\in[0,t]$. To prove the other direction, let $r^*=\arg\max_{s\in[0,t]}\{A_{0,s}-P^0_{0,s}\}$. Pick smallest $r_{p*}\in\mathcal{T}_{P^0}$ with $r_{p*}\geq r^*$. Clearly, $P^0_{r^*,r^-_{p*}}=0$ by definition and $A_{r^*,r^-_{p*}}\geq0$.
\end{Proof}
 
Next, let $C_\alpha$ denote the number of adversarial arrivals starting from time $0$ before the first honest block arrives. Clearly, $C_\alpha$ is a geometric distribution counting the failures where the success probability is $\alpha$. Let $r_{p_i}\in \mathcal{T}_{P^0}$ denote the $i$th pacer of $P^0$, then, $A_{0,r_{p_i}^-}-P_{0,r_{p_i}^-}$ is identically distributed as $C_{\alpha,i}+\sum_{l=1}^{i-1}(C_{\alpha,l}+C_{\Delta,l})-(i-1)$ where $C_{\alpha,l}$ and $C_{\Delta,l}$ are i.i.d.~as $C_{\alpha}$ and $C_{\Delta}$, respectively. Combining this with Lemma~\ref{pacer-end-maximizes}, we obtain
\begin{align}
    P&(L_{t}\geq x)\leq P(\sup_{t\geq q\geq0}\{C_\Delta + \Tilde{A}_{0,q}+\Tilde{P}^0_{0,q}\} \geq x)\\
    &=P(\sup_{ i\geq 0}\{C_{\alpha}+C_{\Delta} +\sum_{l=1}^{i}(C_{\alpha,l}+C_{\Delta,l}-1)\}\geq x)\\
    &=P(1+Z+ \sup_{ i\geq 0}\{\sum_{l=1}^{i}Z_{l}\}\geq x), \label{eqn-just-before-lindley}
\end{align}
where $Z=C_{\alpha}+C_{\Delta}-1$ and $Z_{l}$ i.i.d.~with $Z$. 

Let us define the Lindley process $W_{n+1}=(W_n+Z_{n+1})^{+}$ with $W_{0}=0$, where $(f)^+=\max(f,0)$. Then, $\mathbf{w}_{n}=\mathbf{w}_{0}P^{n}$, where $\mathbf{w}_{n}$ is a column vector representing distribution of $W_n$ and $P$ is a stochastic matrix of $M/G/1$ type
\begin{align}
    P=\begin{bmatrix}
        z_{-1}+z_{0} &  z_{1}  & z_{2} & z_{3} & z_{4} & \ldots\\
        z_{-1} & z_{0} & z_{1}  & z_{2} & z_{3} & \ldots\\
        0          & z_{-1} & z_{0} & z_{1}  & z_{2} & \ldots\\
        0      & 0    & z_{-1} & z_{0} & z_{1}    & \ldots\\
        \vdots & \vdots & \vdots & \vdots & \vdots & \ddots \\
    \end{bmatrix}, \label{pmatrix}
\end{align}
where $z_{i}=P(Z_{l}=i)$. Using Ramaswami's formula\cite{ramaswami-mg1}, we find the steady state distribution of $P$, denoted by $\Pi=\{ \pi_i \}$, which is the limiting distribution of the Lindley process defined above.

\begin{lemma}\label{ramaswami-lemma}
    {\normalfont \textbf{(Ramaswami\cite{ramaswami-mg1})}} The steady-state distribution $\Pi$ can be found recursively by, 
    \begin{align}
        \pi_i&=\frac{\sum_{j=0}^{i-1}\pi_j P(Z\geq i-j)}{z_{-1}}, \quad i\geq1, 
        \label{ramaswami-form}
    \end{align}
    where
    $\pi_0=\frac{-\mathbb{E}\left[Z\right]}{z_{-1}}$.
\end{lemma}

\begin{Proof}
    We apply the Ramaswami's formula 
    \begin{align}
        \pi_i&=\frac{\pi_0 \Bar{b}_i+\sum_{j=1}^{i-1}\pi_j\Bar{a}_{i+1-j}}{1-\Bar{a}_1}, \quad i\geq1, 
        \label{ramaswami-form1}
    \end{align}
    with $\Bar{a}_{i}=\sum_{j\geq i}P_{2,j+1}=P(Z\geq i-1)$ and $\Bar{b}_{i}=\sum_{j\geq i}P_{1,j+1}=P(Z\geq i)$.
    Summing \eqref{ramaswami-form1} for $i\geq1$, yields
    \begin{align}
        (1-\Bar{a}_1)\sum_{i\geq1}\pi_i&=\sum_{i\geq1}\left(\pi_0 \Bar{b}_i+\sum_{j=1}^{i-1}\pi_j\Bar{a}_{i+1-j}\right)\\
        &=\pi_0\sum_{i\geq1}\Bar{b}_i+\sum_{i\geq2}\sum_{j=1}^{i-1}\pi_j\Bar{a}_{i+1-j}\\
        &=\pi_0\sum_{i\geq1}\Bar{b}_i+\sum_{j\geq1}\pi_j\sum_{i\geq2}\Bar{a}_{i}.
    \end{align}
    Reordering the terms, we obtain
    \begin{align}
        (1-\pi_0)&=\pi_0\sum_{i\geq 1}\Bar{b}_i+(1-\pi_0)\sum_{i\geq 1}\Bar{a}_i\\
        &=\pi_0\mathbb{E}\left[Z^+\right]+(1-\pi_0)\mathbb{E}\left[Z+1\right],
    \end{align}
    which in turn results in $\pi_0=\frac{-\mathbb{E}\left[Z\right]}{z_{-1}}$. Expressing $\Bar{a}_i$ and $\Bar{b}_i$ in terms of the distribution of $Z$ and combining the terms in the numerator concludes the proof. Note that we need $1>\mathbb{E}\left[C_{\alpha}+C_{\Delta}\right]$, i.e., the number of adversarial arrivals before the honest arrival and during its delay should be on average less than $1$. In other words, the rate of honest arrivals under the maximum delay strategy should be greater than the rate of adversarial arrivals.
\end{Proof}

\begin{lemma}
    \label{asmussen-lemma}{\normalfont \textbf{(Asmussen\cite[Corollary 3.2]{asmussen-ruin})}} The Lindley process $W_{n+1}=(W_n+Z_{n+1})^{+}$ with $W_{0}=0$ where $(f)^+=\max(f,0)$ converges in distribution to $\sup_{ i\geq 0}\{\sum_{l=1}^{i}Z_{l}\}$.
\end{lemma}

\begin{corollary}
    The pre-mining gain is bounded by $\Bar{L}$ where $P(\Bar{L}=0)=-\mathbb{E}\left[Z\right]$, $P(\Bar{L}=1)=\pi_0+\mathbb{E}\left[Z\right]$ and $P(\Bar{L}=i)=\pi_{i-1}$ for $i\geq 2$.
\end{corollary}
\addtolength{\topmargin}{+0.01in}
\begin{Proof}
    By Lemma~\ref{ramaswami-lemma} and Lemma~\ref{asmussen-lemma}, $\sup_{ i\geq 0}\{\sum_{l=1}^{i}Z_{l}\}$ has the same distribution as $\Pi$. We rewrite \eqref{eqn-just-before-lindley} to obtain
    \begin{align}
        P(L\geq x)&\leq P(1+Z+ \Pi \geq x)\\
        &=P(\Bar{L}\geq x),
    \end{align}
    where $\Bar{L}=1+Z+ \Pi$. Thus, for all $i\geq 0$, we have,
    \begin{align}
        P(\Bar{L}=i) &=\sum_{j=0}^{i}\pi_j z_{i-j-1}.\label{lead_conv_sum}
    \end{align}
Since $\Pi P=\Pi$, for $i\geq 1$, we also have,
    \begin{align}
        \pi_i &=\sum_{j=0}^{i+1}\pi_j z_{i-j}.\label{pi_conv_sum}
    \end{align}
The result follows from \eqref{lead_conv_sum} and \eqref{pi_conv_sum}. 
\end{Proof}

We note that for bounded delay model, $\Bar{L}$ is the time domain equivalent of the probability generating function provided in \cite[Lemma 5]{cao2023tradeoff}. Notice, with the method we provide, it is much easier to evaluate the lead as it does not require exhaustive computation of derivatives. When transaction $tx$ arrives at the system at time $\tau_0$, it will be included in the next honest block. Note that, the time $tx$ arrives might correspond to a delay interval of a jumper that arrived just before $\tau_0$, in which case, the adversary can choose to fork (and hence, void) the first honest block containing $tx$, which allows the adversary an additional time for increasing the lead before a jumper block containing $tx$ is mined. The extra advantage gained in this scenario by the adversary is already accounted for in the lead distribution computed above, since we allow $q\geq 0$ in \eqref{lead_bound_tildes} and consider the best case scenario for the adversary, i.e., we consider the lead just before the arrival of a new jumper ($s^-$ in Lemma~\ref{pacer-end-maximizes}) and do not count the negative effects of this new jumper on the lead. Thus, we can simply assume that $\tau_0$ is a time where the first honest block to be mined after $\tau_0$ is going to be a jumper that contains $tx$.

Since the arrival of $tx$ implies that the adversary can deploy balancing attack strategy, during $[\tau_0, \tau_k]$ we assume that all honest blocks arriving during the delay $\Delta_h$ of a jumper block at height $h$ are rigged (converted to adversarial blocks), which allows us to use \cite[Thm.~3]{guo-btc-sec-lat} that states if any attack succeeds in violating a transactions safety, then so does the private attack. 

Let $\Bar{C}_\Delta$ denote the random variable that represents the total number of blocks mined during a delay interval $\Delta_h$. Then, the number of adversarial blocks mined (including the rigged blocks) between the publication of two jumper blocks, which are delayed by maximally allowed time, are i.i.d.~with $C_{\alpha} + \Bar{C}_{\Delta}$. This in turn implies that the number of adversarial blocks mined during the confirmation interval is bounded by the sum of $k$ i.i.d.~random variables $C_{\alpha} + \Bar{C}_{\Delta}$, denoted as $\Bar{S}_k$.

At $\tau_k$, $tx$ is confirmed in all honest views and the adversary can undo the confirmed $tx$ if the sum of the lead and the number of adversarial blocks mined during the confirmation interval exceeds $k$. Else, the adversary has a deficit $D$ of at least $k-\Bar{S}_k-\Bar{L}$ that it has to make up relative to the honest chain after $\tau_k$. But since the block containing $tx$ is confirmed in all honest views at $\tau_k$, the private chain containing $tx'$ can be only extended by the adversary until it catches the honest chain containing $tx$, hence we do not need to empower the adversary with rigged blocks after $\tau_k$. 

Before the violation happens, for any time $d\geq\tau_k$, during the post confirmation race, the growth in the private chain is bounded above by $A_{\tau_k,d}$, whereas the growth in the honest chain containing $tx$ is bounded below by $J_{\tau_k,d}=P^{\tau_k}_{\tau_k,d}$. Further, for any jumper mining time $t_j$ at some height $h_j$ after $\tau_k$, the adversary can delay the publication of that jumper by $\Delta_{h_{j}}$ to gain some extra time to catch the honest chain. Let $\Bar{d}=d$ if $d\notin [t_j, t_j+\Delta_{h_j}]$ for any $j$ and $\Bar{d}=t_j+\Delta_{h_j}$ if $d\in [t_j, t_j+\Delta_{h_j}]$ for some $j$. Then, the discussion above implies that the distribution of the maximum deficit adversary can make up during the post confirmation race, denoted as $M$, is stochastically bounded by
\begin{align}
    P(M\geq x)&= P(\sup_{d\geq \tau_k}\{A_{\tau_k,\Bar{d}}-J_{\tau_k,d^-}\} \geq x)\label{deficit-eqn}\\
    &= P(\sup_{d\geq 0}\{\Tilde{A}_{0,\Bar{d}}-\Tilde{P}^{0}_{0,d^-}\} \geq x)\\
    &= P(\sup_{d\in \mathcal{T}_{\Tilde{P}^0}}\{\Tilde{A}_{0,d}+C_{\Delta}-\Tilde{P}^{0}_{0,d^-}\} \geq x)\\
    &= P(1+Z_0+ \sup_{ i\geq 0}\{\sum_{l=1}^{i}Z_{l}\}\geq x)\\
    &= P(1+Z+ \Pi \geq x)\\
    &=P(\Bar{M}\geq x), \label{deficit-2-eqn}
\end{align}
where we apply the same reasoning and the steps as we did in bounding the lead $L$. Notice that, the bound we give for pre-mining gain and post confirmation race are distributed i.i.d.~with $\Bar{L}$. We can combine the analyses of the three phases: pre-mining gain, confirmation interval and post confirmation race to bound the safety violation probability $p(\alpha,\mu_m,\Delta_h,k)$.

\begin{theorem}
    Given mining rate $\mu_m$, honest fraction $\alpha$, delay distributions of $\Delta_h$ and confirmation depth $k$, a confirmed transaction cannot be discarded with probability greater than,
    \begin{align}
     p(\alpha,\mu_m,\Delta_h,k)\leq\Bar{p}(\alpha,\mu_m,\Delta_h,k)=P(\Bar{L}+\Bar{S}_k+\Bar{M}\geq k). \label{safety-vio}
    \end{align}
\end{theorem}

\subsection{The Case of $b_0>0$}

We start by considering the effect of $b_0>0$ on the lead distribution. Clearly, Lemma~\ref{jumper-start-maximizes} is still valid for any $b_0$ and adversary should delay the publication of any jumper block after the starting jumper block by maximally allowed delay time to maximize the lead. Let $t_j$ denote the mining time of the jumper block at height $h_j$. After $t_j+\Delta_{h_j}$, all honest miners are aware of the jumper block at height $h_j$. If $b_0$ HP transactions did not arrive in the interval $[t_j,t_j+\Delta_{h_j}]$, the honest miners cannot start mining on height $h_j+1$ at $t_j+\Delta_{h_j}$ despite being aware of the block at height $h_j$ and have to wait until there are $b_0$ HP transactions in their mempool. Further, assume that another honest block was mined at $t'_j\in (t_j,t_j+\Delta_{h_j}]$ creating a fork at height $h_j$, we call such a block as forked block. Since adversary has to publish the jumper on height $h_j$ at $t_j+\Delta_{h_j}$, it does not lose anything by publishing the honest block that was mined at $t'_j$ together with the jumper block. On the other hand, any HP transaction that arrived in the interval $[t_j,t'_j]$ can be assumed to be included in the block at height $h_j$ since ties are broken in adversary's favor, which further delays the start of the honest mining on the height $h_j+1$. Thus, the best adversary can do in order to delay the mining on the height $h_j+1$, is to delay the publication of the jumper at $h_j$ by $\Delta_{h_j}$, publish all honest blocks at height $h_j$ at time $t_j+\Delta_{h_j}$ and let the honest miners mine on the tip of the honest block mined at the largest honest mining time $t'_j\in [t_j,t_j+\Delta_{h_j}]$. The effective idle time honest miners spend after the mining of a jumper block can be expressed as $\max(\Delta_{h_j}, t'_j-t_j+E_{b_0}(\lambda_h))$ for the jumper at height $h_j$, where $E_{k}(\lambda)$ denotes Erlang-$k$ distribution and $t'_j$ is the mining time of the last honest block mined during $[t_j,t_j+\Delta_{h_j}]$. The length of this idle time can be further upper bounded by $\Delta_{h_j}+E_{b_0}(\lambda_h)$. By denoting the number of adversarial blocks mined during $E_{b_0}(\lambda_h)$ as $C_{b_0}$, we can repeat the pre-mining results found previously for $b_0=0$ by simply replacing definition of $Z$ with $Z^{(b_0)}=C_{\alpha}+C_{\Delta}+C_{b_0}-1$ and denote the bound on the lead as $\Bar{L}^{(b_0)}$.

Similar to the argument for the case $b_0=0$, here we can assume that $\tau_0$ is a time where all honest miners are active and the first honest block after $\tau_0$ is a jumper block that contains $tx$ since we consider the best case scenario while bounding the lead. Since adversary can deploy a balancing attack strategy, during $[\tau_0,\tau_k]$, we assume that all honest blocks are rigged during the delay of a jumper block. Further, there are sample paths of arrivals for HP transactions and blocks such that adversary might benefit from a forked block instead of a rigged block since rigged blocks imply that all HP transactions that arrive during the delay interval are added to the mempool whereas a forked block implies some of the HP transactions are already mined, which delays the honest mining process. By assuming that no HP transaction arrives during delay intervals, we give an additional power to the adversary to avoid such situations, which allows us to use \cite[Thm.~3]{guo-btc-sec-lat} again. Thus, the honest blocks are rigged during a delay interval of a jumper block mined during the confirmation interval and honest miners are inactive from the end of the delay interval until $b_0$ HP transactions arrive at the system. This in turn implies that the number of adversarial blocks mined during the confirmation interval is bounded by the sum of $\Bar{S}_k$ and $k-1$ i.i.d.~random variables $C_{b_0}$, denoted by $S^{(b_0)}_k$. 

After $\tau_k$, the maximum deficit adversary can make up relative to the honest chain is upper bounded using the same steps as from \eqref{deficit-eqn} to \eqref{deficit-2-eqn}, where we replace $Z$ with $Z^{(b_0)}$. Thus, the bound we give for pre-mining gain and post confirmation race with $b_0$ condition are distributed i.i.d.~with $\Bar{L}^{(b_0)}$.

\section{Lower Bound}

For the case $b_0=0$, we simply consider the adversarial strategy that tries to build a lead starting from the genesis block and delays each jumper by maximal delay allowed. The lead $L_t$ can be bounded below with the Lindley process $W_{n+1}=(W_n+Z_{n+1})^{+}$ where $W_{0}=(C_\alpha-1)^+$ and assuming the time $tx$ arrives is large enough, the lead can be bounded as
\begin{align}
    P(L_{t}\leq x)&\leq P(\Pi+C_\Delta \leq x)
\end{align}

Notice the difference between upper and lower bound on $L_t$: While upper bounding, we consider the best case scenario and assume the adversary starts mining right after a jumper that gives it an advantage of $C_\Delta$ on top of $\Pi$ and the race ends just before the mining time of a jumper thus adversary gains another advantage of $C_\alpha$. While lower bounding, the pre-mining race starts on equal terms right after the genesis block and ends just after the mining of a jumper honest block, which is the lowest point of the lead, after which adversary delays the publication for $\Delta_h$ hence only $C_\Delta$ term is present in the lower bound. For certain parameter regimes, there are several ways to further tighten the lower bound  on the lead such as alternating renewal theorem or considering a simple random walk without any delay strategy during pre-mining phase, which we skip due to space limitations.

After $\tau_0$, we can assume the first honest block to be mined contains $tx$ and is going to be a jumper, which in turn means we can bound the number of adversarial blocks during confirmation interval as sum of $k$ i.i.d.~random variables $C_\alpha+C_\Delta$ denoted by $S_k$. The deficit adversary can make up during post confirmation race was already calculated as $M$. If $b_0>0$, the lower bound can be further tightened by replacing $\Delta_h$ with $\max(\Delta_h,E_{b_0}(\lambda_h))$ and redefining $C_{\Delta}$ accordingly.

\begin{figure}[t]
    \centerline{\includegraphics[width=0.8\columnwidth]{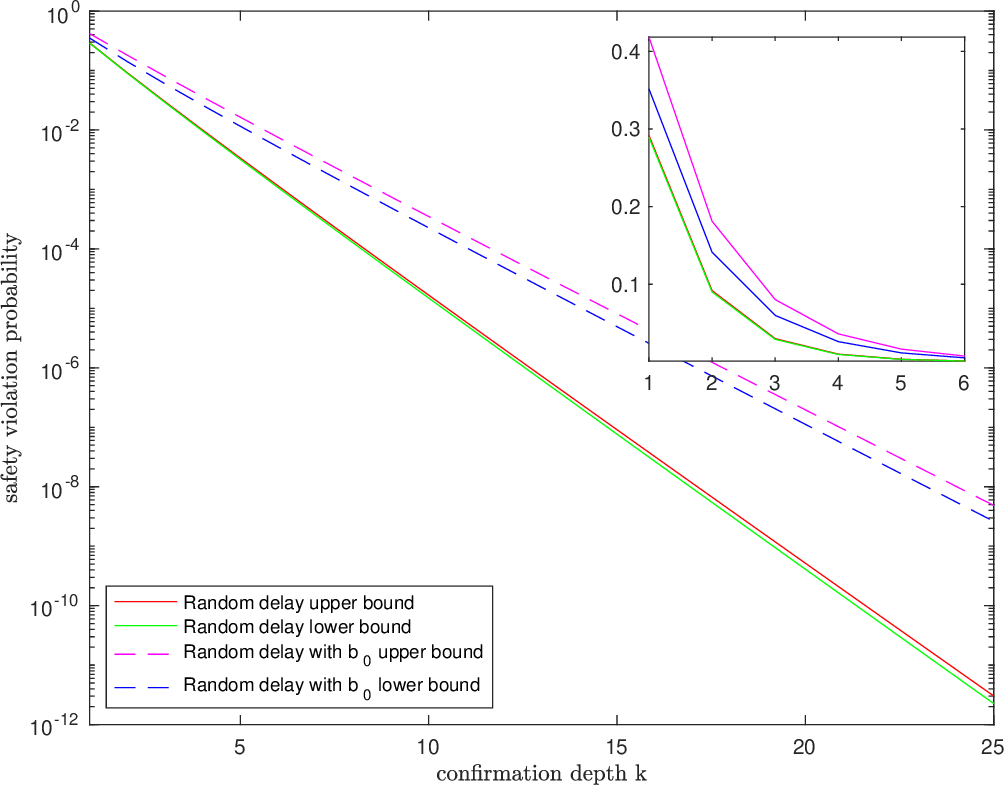}}
    \caption{Safety violation for random delay model, $\alpha=0.9$.}
    \label{btc-rand-delay-fig}
\end{figure}

\section{Numerical Results}

The security-latency results provided in \cite{our-sec-lat-extended,our-queue-sec-ext-version} are improved with the methods provided here. As the state-of-the-art results are already tight and visually indistinguishable, we skip plotting them here for bounded and exponential delay models. However, unlike the earlier results, our method is applicable to any random delay distribution, hence we pick a more realistic right-skewed delay distribution with $\Delta_h \sim E_2(1)$, where the $90$th percentile block propagation delay is around $4$ seconds and fits well on empirical pdf provided in \cite{DSN-Bitcoin-Monitoring} for BTC with $\mu_m=1/600$. \figref{btc-rand-delay-fig} displays upper and lower bounds for $b_0=0$ as well as $b_0=100$ with $\lambda_h=1/5$. The stability condition and our assumptions on mempool can be expressed $\frac{b}{\lambda_h}\gg \frac{1}{\mu_m}+E[\Delta]$ since we assume HP mempool is not full. Further, for $b_0$ condition to have a considerable impact on security we need $\frac{b_0}{\lambda_h}\gg E[\Delta]$. The choice of $b_0$ and $\lambda_h$ here can be further interpreted with real-world data of transaction fees in order for classifications of transactions which we skip here due to space limitations. Here, for the $6$-block confirmation rule of BTC, the safety violation probability's upper and lower bounds are narrowed to $0.00115$ and $0.00108$, respectively, ($0.0075$ and $0.0055$ with $b_0=100$).

\newpage
\bibliographystyle{ieeetr}
\bibliography{lib}
\end{document}